\documentclass[reprint,twocolumn,floatfix,aps,longbibliography]{revtex4-2}
\usepackage[T1]{fontenc}
\usepackage{amsmath,amssymb}
\usepackage{epsfig}
\usepackage[utf8]{inputenc}
\usepackage{amsmath,amsfonts}
\usepackage{graphicx}
\usepackage{color}
\usepackage{hyperref}
\usepackage{turnstile}
\usepackage{relsize}
\usepackage[toc,page]{appendix}
\usepackage{pstricks-add}
\usepackage{relsize}
\usepackage{mathtools}
\usepackage{physics}
\usepackage{tikz}

\usetikzlibrary{positioning,shapes.arrows,decorations.pathreplacing,intersections,decorations.markings}
\usepgfmodule{nonlineartransformations}
\usepgflibrary{curvilinear}
\usetikzlibrary{spy,calc}

\usepackage{tikz-3dplot}
\usepgflibrary{fpu}
\newcommand{\PgfmathsetmacroFPU}[2]{\begingroup
    \pgfkeys{/pgf/fpu,/pgf/fpu/output format=fixed}%
    \pgfmathsetmacro{#1}{#2}%
    \pgfmathsmuggle#1\endgroup}%
\makeatletter
\def\fancyspheretransformation{
    \PgfmathsetmacroFPU{\XYTwo}{\pgf@x*\pgf@x+\pgf@y*\pgf@y}%
    \PgfmathsetmacroFPU{\zvalue}{(1.2*\XYTwo)/(\XYTwo/\spradius+4*\spradius)}%
    \PgfmathsetmacroFPU{\xvalue}{(2*\spradius+\zvalue)/(2*\spradius)*\pgf@x}%
    \PgfmathsetmacroFPU{\yvalue}{(2*\spradius+\zvalue)/(2*\spradius)*\pgf@y}%
    \PgfmathsetmacroFPU{\myx}{cos(\tdplotmainphi)*\xvalue+sin(\tdplotmainphi)*\yvalue}%
    \PgfmathsetmacroFPU{\myy}{-cos(\tdplotmaintheta)*sin(\tdplotmainphi)*\xvalue+cos(\tdplotmaintheta)*cos(\tdplotmainphi)*\yvalue-sin(\tdplotmaintheta)*\zvalue}%
    \pgf@y=\myy pt%
    \pgf@x=\myx pt%
}
\makeatother

\usepackage{xifthen}
\usepackage{import}
\usepackage{transparent}

\def\Bf#1{\mbox{\boldmath $#1$}}
\begin{document}

\title{Projection Optimization Method for Open-Dissipative Quantum Fluids\\ and its Application to a Single Vortex in a Photon Bose--Einstein Condensate}

\author{Joshua Krau{\ss}$^1$, Marcos Alberto Gon\c{c}alves dos Santos Filho$^{1,2}$, Francisco Ednilson Alves dos Santos$^2$, and Axel Pelster$^1$}
\affiliation{$^1$Physics Department and Research Center OPTIMAS, University Kaiserslautern-Landau,
Erwin-Schrödinger Stra{\ss}e 46, 67663 Kaiserslautern, Germany\\ $^2$Departamento de Física, Universidade Federal de S\~{a}o Carlos, S\~{a}o Carlos, S\~{a}o Paulo, 13565-905, Brazil\\
{\tt jkrauss@rptu.de, malberto@df.ufscar.br, santos@ufscar.com, axel.pelster@rptu.de}}

\begin{abstract}
Open dissipative systems of quantum fluids have been well studied numerically. In view of a complementary analytical description we extend here the variational optimization method for Bose--Einstein condensates of closed systems to open-dissipative condensates. The resulting projection optimization method is applied
to a complex Gross--Pitaevskii equation, which models a photon Bose--Einstein condensate. Together with known methods from hydrodynamics we obtain an approximate vortex solution, which depends on the respective open system parameters and has the same properties as obtained numerically in the literature.
\end{abstract}

\maketitle
\section{\label{sec:Intro} Introduction}
Atomic or molecular Bose--Einstein condensates represent closed quantum many-body systems as their magneto-optical or optical trapping isolates them perfectly from the laboratory environment.
Their dynamics are generically described by the time-dependent Gross-Pitaevksii equation \cite{Gross,Pitaevskii}.
Its solutions in the absence or presence of  vortices have been investigated by numerous studies either numerically or analytically.
In the former case, modern programming languages supporting open multiprocessing are now available, significantly reducing execution time on multicore processors (see, e.g., Refs.~\cite{Antun1,Antun2}).
The latter case is typically based
on a variational optimization method, which relies on the existence of an underlying action. With a suitably chosen trial condensate wave function
the corresponding variational parameters are fixed by applying the Hamilton principle, which qualitatively captures the physics to be investigated.
Thus, the spatial degrees of freedom are eliminated and the quest for solving the
Gross--Pitaevskii equation is reduced to solving coupled nonlinear
ordinary differential equations for the time dependence of the variational parameters.
In this way, for instance, the low-lying excitation modes and frequencies of Bose--Einstein condensates can successfully be determined \cite{Zoller1,Zoller2,Fetter}.

Bose–-Einstein condensates can occur not only in closed but also in open-dissipative quantum many-body systems. Modern prime examples are provided by
quasi-equilibrium magnons at room temperature under pumping \cite{Hillebrands} and
exciton-polariton condensates in a gallium arsenide microcavity \cite{Dang}. Nowadays also
photon condensates are relevant, which have been observed in dye-filled microcavities \cite{Weitz2010} and, quite recently, also in a vertical-cavity surface-emitting laser \cite{vcsel3,vcsel1,vcsel2}.
Another upcoming platform is a hybrid atom-optomechanical system in which a single mechanical mode of a
nanomembrane in a cavity is optically coupled to a far distant atomic Bose--Einstein condensate residing in the
potential of the out-coupled standing wave of the cavity light \cite{Treutlein}. As for such open-dissipative systems
both the energy and the particle number are not conserved quantities, an action for deriving the underlying equations of motion is not readily available.
For example, the variational optimization method outlined above for closed systems cannot be straightforwardly applied to open-dissipative systems. It would require a careful adaptation within the corresponding Schwinger–Keldysh action framework, see the review in Ref.~\cite{Diehl} and references therein.
A priori it is unclear which variational ansatz would be reasonable for both the classical and the quantum field on the microscopic level.
This problem can be circumvented by considering the equations of motion for the cumulants of the condensate wave function \cite{Kubo,Leymann,Radonjic,Nagler}.
They can be self-consistently evaluated with a corresponding ansatz for the condensate wave function, where its cumulants represent the variational parameters to be determined.
Most common is a Gaussian approximation neglecting all the cumulants of higher than second order.
Thus, one focuses on calculating the particle number, the centre-of-mass and the width as the zeroth, first, and second cumulant of the condensate wave function, respectively.
This allows, for instance, to successfully  describe
the nonequilibrium quantum phase transition in a hybrid atom-optomechanical system \cite{Thorwart,Thorwart2019} and to determine
the collective modes of a photon Bose--Einstein condensate with thermo-optic interaction \cite{Stein}. Furthermore, it should be noted that applying the  cumulant optimization method to closed systems
yields the same results as the variational optimization method.

However, the  cumulant optimization method has a limited range of applicability. For practical reasons it can only be applied to physical situations,
where the condensate wave function consists of a finite number of cumulants. For instance, it can not be applied to describe a vortex in an open-dissipative system like a photon condensate,
which is of fundamental physical interest.
On the one hand, the effective interaction strength for photon condensates is so small \cite{Weitz2010,Stein,Klaers2011,Stein2022,Stein2023b} that
one would expect from the standard Gross--Pitaevskii theory such a large healing length
that no vortex could fit into a finite-sized system. But, surprisingly, the interplay of pumping and nonlinear losses can nevertheless lead to a finite vortex core size \cite{Wouters1}.
On the other hand, such a vortex turns out to behave differently than in the standard Gross--Pitaevskii theory for closed systems. Due to the open-dissipative nature of such a quantum fluid its velocity field  has not only
an incompressible tangential component but also a compressible radial component, yielding a spiral vortex shape \cite{Wouters1, Wouters2}.
The radial velocity component implies that the vortex does not preserve the particle number, thus it acts as a particle source.
Because of the reasons mentioned it would be useful to have also in such open-dissipative systems an optimization method available which is capable of determining the properties
of those spiral vortices in a quantitative way.

To this end we proceed in the rest of the paper as follows. In Section \ref{sec:Proj} we introduce a projection optimization method for open-dissipative systems, which contains the cumulant optimization method as well as the variational optimization method as a special case. Afterwards,
this is applied to a complex Gross--Pitaevskii equation \cite{Keeling2008} in Section \ref{sec:Appl} , where the interplay between pumping and nonlinear losses gives rise to a spiral vortex.
The obtained approximate analytical results are then compared in Section \ref{numerics} with a numerical solution of the underlying complex Gross--Pitaevskii equation. In particular, we focus the discussion on the radial velocity as it stems from the openness of the considered system.
\section{\label{sec:Proj} Projection Optimization Method}
Optimization methods are essential in finding approximate analytical solutions. Especially in arbitrary systems, where no action based principle is applicable, finding a solution can be quite complicated. Therefore, we generalize in Subsection \ref{general} the variational optimization method to arbitrary systems by invoking a projection optimization method. Subsequently Subsection \ref{geometry}
discusses how our approach can be visualized and geometrically interpreted. 
\subsection{General Formulation}\label{general}
In the following we formulate the projection optimization method from a general point of view. Let us
consider an arbitrary system described by the field ${\bf \Psi} =\left(\Psi^1, \ldots , \Psi^N\right)$, where $\Psi^i$ denotes an element of the underlying Hilbert space $\mathcal{H}=\left(H,\langle\bullet,\bullet\rangle\right)$, which consists of a $\mathbb{C}$-vector space $H$ and a scalar product $\langle\bullet,\bullet\rangle$.
In addition, we assume that the field fulfills the equation of motion
\begin{eqnarray}
\label{EOM}
\textrm{EOM}[{\bf\Psi}^{*}, {\bf \Psi}] = 0.
\end{eqnarray}
The aim is now to find an approximate solution of (\ref{EOM}) by using a suitable trial field ${\Bf \psi}({\Bf \alpha})$, which depends on a set of trial parameters ${\Bf \alpha}$ with $\alpha^i\in\mathbb{R}$, according to ${\bf \Psi} \approx {\Bf \psi}({\Bf \alpha})$. To this end we use the scalar product in order to
project the equation of motion for the trial field onto a parameter manifold, which is spanned by the trial parameters. The latter can be determined by solving the set of coupled algebraic equations
\begin{equation}
\label{variational-projection}
    \left\langle{
    \textrm{EOM}^{*}[{\Bf \psi}^{*},{\Bf \psi}]},{\frac{\partial{\Bf \psi}^{*}}{\partial{\alpha^{i}}}}\right\rangle + \left\langle{
    \textrm{EOM}[{\Bf \psi}^{*},{\Bf \psi}]},{\frac{\partial{\Bf \psi}}{\partial{\alpha^{i}}}}\right\rangle = {\bf 0},
\end{equation}
where $\partial{\Bf \psi}/\partial{\alpha^{i}} = \left(\partial\psi^1/\partial\alpha^i, \ldots , \partial\psi^N/\partial\alpha^i\right)$.
Note that the method can be spatio-temporally generalized by using ${\bf \alpha}={\bf \alpha}({\bf x},t)\in\mathbb{C}$ and by exchanging partial derivatives with
functional derivatives. With this
(\ref{variational-projection}) will become in general a set of partial differential equations instead of algebraic equations.

This projection optimization method can be motivated heuristically as it reduces to the variational optimization method for closed systems \cite{Zoller1,Zoller2}.
In the following we restrict ourselves for the sake of simplicity to the case $N=1$, as the generalization for $N>1$ is straight-forward. Furthermore, we assume that some action ${\cal A}[\Psi^*, \Psi]$ exists, which can be approximated with a trial ansatz for the field as mentioned above according to ${\cal A} ({\Bf \alpha}) \approx {\cal A}[\psi^* ( \alpha), \psi(\alpha)]$. Then we specify the general Hilbert space $\mathcal{H}$ to be the space $L^2$ over $\mathbb{C}$ with the inner product
\begin{equation}
    \label{eq:inner-product}
    \langle{f},{g}\rangle \equiv \int{f^{*}({\bf x})g({\bf x})}\dd[D]{x}\, .
\end{equation}
An extremization of the approximated action with respect to the trial parameters thus yields the variational equations
\begin{eqnarray}
\frac{\delta {\cal A}}{\delta \alpha^i}=
\int \left( \frac{\delta {\cal A}}{\delta \psi^*} \, \frac{\partial \psi^*}{\partial \alpha^i} +
 \frac{\delta {\cal A}}{\delta \psi} \, \frac{\partial \psi}{\partial \alpha^i} \right)\mathrm{d}^Dx  = 0\,.
\end{eqnarray}
They can be rewritten using~\eqref{eq:inner-product} as
\begin{equation}
   \left\langle{\frac{\delta {\cal A}}{\delta \psi}},{\frac{\partial{\psi^*}}{\partial{\alpha^{i}}}}\right\rangle + \left\langle{\frac{\delta {\cal A}}{\delta \psi^*}},{\frac{\partial{\psi}}{\partial{\alpha^{i}}}}\right\rangle
    = 0\, ,
\end{equation}
which corresponds to \eqref{variational-projection} for closed systems using the identification
\begin{eqnarray}
\label{variation}
\frac{\delta{\cal A}}{\delta \Psi^*}=\textrm{EOM} [\Psi^*, \Psi]\,.
\end{eqnarray}
Thus, in case (\ref{variation}) holds, the projection optimization method is, indeed, equivalent to the variational optimization method for closed systems. Moreover, there exist many special cases of the projection optimization method in the literature. For instance, the equivalence of the method proposed in Ref.~\cite{Gammal2001_variation} for determining the dynamics of a trapped Bose-Einstein condensate can straight forwardly be shown. Furthermore, the cumulant optimization method proposed for describing the nonequilibrium quantm phase transition in a hybrid atom-optomechanical system
\cite{Thorwart,Thorwart2019} and for calculating
the collective modes of a photon Bose--Einstein condensate with thermo-optic interaction \cite{Stein} represents a special case of the projection optimization method.
In addition, beyond mean-field calculations of steady-state properties of open-dissipative systems governed by a Lindblad master equation in 
Ref.~\cite{Weimer} turns out to also be equivalent by identifying appropriately the underlying scalar product.
\subsection{Geometrical Interpretation} \label{geometry}
We show now that the projection optimization method can also be visualized and geometrically interpreted. To this end we consider for the sake of simplicity the example of a system of first-order ordinary differential
equations describing the time evolution of an $N$-dimensional vector ${\bf X}=\left(X^1, \ldots , X^N\right)$ according to
\begin{eqnarray}
\label{dynamics1}
\textrm{EOM} ({\bf X}) = \dot{\bf X} - {\bf F} ({\bf X})={\bf 0} \, ,
\end{eqnarray}
where ${\bf F}({\bf X})=(F^1({\bf X}), \ldots , F^N({\bf X}))$ represents some underlying vector field. Let us assume that this dynamics is approximately described by $N$ trial functions ${\bf x}({\Bf \alpha})$, which depend on a set of $M$ trial
parameters ${\Bf \alpha}=(\alpha^1,\ldots,\alpha^M)$. Note that we don't restrict the choice of $N$ and $M$.

By identifying ${\bf X}\approx {\bf x}({\Bf \alpha})$ we aim at projecting approximately
the original time evolution of ${\bf X}(t)$ to a corresponding one ${\Bf \alpha}(t)$ for the trial parameters. Thus, the goal is to map the set of differential equations (\ref{dynamics1})
to another one
\begin{eqnarray}
\label{dynamics2}
\textrm{EOM}({\Bf \alpha})=\dot{\Bf \alpha} - {\bf f} ({\Bf \alpha}) ={\bf 0}
\end{eqnarray}
with a yet to be determined vector field ${\bf f} ({\Bf \alpha})=(f^1({\Bf \alpha}), \ldots, f^M({\Bf \alpha}))$.

\begin{figure}[t]
    \centering
\includegraphics[width=\columnwidth]{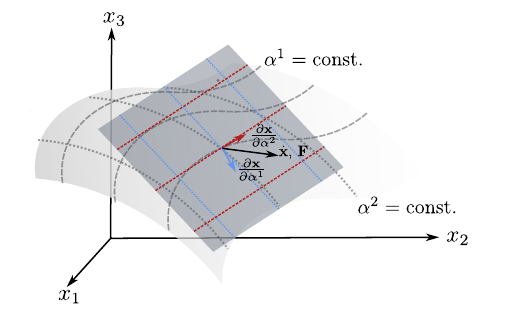}
    \caption{Illustration of projection optimization method for $N=3$ and $M=2$. Approximative ansatz ${\bf X}\approx {\bf x}({\Bf \alpha})$ implies that velocity $\dot{\bf x}$ and vector field ${\bf F}({\bf x})$ roughly lie in tangent plane of manifold ${\bf x}({\Bf \alpha})$ at point ${\bf x}$ spanned by tangent vectors $\partial {\bf x}/\partial \alpha^1$ and $\partial {\bf x}/\partial \alpha^2$.}
    \label{projection-illustration}
\end{figure}

According to the projection optimization method the reduction from (\ref{dynamics1}) to (\ref{dynamics2}) is achieved by evaluating
\begin{eqnarray}
\label{dynamics3}
\left\langle \dot{\bf x} - {\bf F} ({\bf x}) , \frac{\partial {\bf x}}{\partial \alpha^i} \right \rangle = 0\, ,\hspace*{0.5cm}i=1,\dots, N\, ,
\end{eqnarray}
where $\langle \bullet,\bullet \rangle$ stands here for the Euclidean scalar product of the $N$-dimensional vector space. This prescription has now a quite intuitive geometrical interpretation.
Namely we recognize that the approximative ansatz ${\bf X}\approx {\bf x}({\Bf \alpha})$ means that the trajectory ${\bf X}(t)$ lies roughly on an $M$-dimensional manifold, whose embedding in the $N$-dimensional
vector space is defined according to ${\bf x}({\Bf \alpha})$. Thus, this manifold is characterized by a
parametrization with respect to the trial parameters ${\Bf \alpha}$, see the illustration in Fig.~\ref{projection-illustration}.
Then the partial derivative $\partial {\bf x}/\partial \alpha^i$ represents a tangent vector to the manifold, which is pointing perpendicular
to the $\alpha^j=\mbox{const}$ line in the tangent plane at point ${\bf x}$. Therefore, Eq.~(\ref{dynamics3}) amounts to projecting the original dynamics (\ref{dynamics1}) with respect to all tangent vectors of the manifold.
This is a natural condition to impose as the approximative ansatz ${\bf X}\approx {\bf x}({\Bf \alpha})$ implies that both the velocity $\dot{\bf x}$ and the original vector field ${\bf F} ({\bf x})$ lie roughly in the
tangent plane of the manifold at point ${\bf x}$ as is depicted in Fig.~\ref{projection-illustration}.
Therefore, they must have the same coefficients with respect to an expansion in the basis of the tangent space, i.e.~the tangent vectors $\partial {\bf x}/\partial \alpha^i$.

Now we evaluate the projection (\ref{dynamics3}) by applying the chain rule and using the Einstein summation convention
\begin{eqnarray}
\label{dynamics4}
\dot{\bf x} = \dot{\alpha}^j\,\frac{\partial {\bf x}}{\partial \alpha^j}\, ,
\end{eqnarray}
which leads to
\begin{eqnarray}
\label{dynamics5}
\dot{\alpha}^j \, \left\langle\frac{\partial {\bf x}}{\partial \alpha^j} , \frac{\partial {\bf x}}{\partial \alpha^i} \right\rangle
= \left\langle {\bf F}\left( {\bf x} \right), \frac{\partial {\bf x}}{\partial \alpha^i} \right\rangle\, . \label{dynamics_projected}
\end{eqnarray}
The scalar product of two tangent vectors on the left-hand side of (\ref{dynamics_projected}) defines the covariant metric of the manifold
\begin{eqnarray}
\label{dynamics6}
g_{ji}({\Bf \alpha}) = \left\langle \frac{\partial {\bf x}({\Bf \alpha})}{\partial \alpha^j} , \frac{\partial {\bf x}({\Bf \alpha})}{\partial \alpha^i} \right\rangle\, ,
\end{eqnarray}
which has the contravariant version $g^{ik}$ as its inverse, according to
\begin{eqnarray}
\label{dynamics9}
g_{ji} g^{ik} = \delta_{j}^{\hspace*{2mm}k}
\end{eqnarray}
with $\delta_{j}^{\hspace*{2mm}k}$ denoting the Kronecker symbol. And the scalar product on the right-hand side of (\ref{dynamics_projected}) defines the projected components of $\mathbf{F}(\mathbf{x})$ on the manifold
\begin{eqnarray}
\label{dynamics7}
f_i({\Bf \alpha}) = \left\langle {\bf F}\left( {\bf x}({\Bf \alpha}) \right), \frac{\partial {\bf x}({\Bf \alpha})}{\partial \alpha^i} \right\rangle ,
\end{eqnarray}
which has its contravariant components given by
\begin{eqnarray}
f^i({\Bf \alpha})=g^{ij}({\Bf \alpha})f_j({\Bf \alpha}).
\end{eqnarray}
Given that {\bf f} is the vector representing the projection of {\bf F} onto the manifold, {\bf f} can be regarded as an approximation of ${\bf F}$ in the sense that they are nearly identical, differing only in the components perpendicular to the manifold, which are omitted in {\bf f}.

The final equation of motion for the parameters ${\Bf \alpha}$ can be obtained by multiplying both sides of (\ref{dynamics_projected}) by $g^{ij}$, thus giving
$\dot{\alpha^i}=f^i({\Bf \alpha})$, which corresponds to (\ref{dynamics2})
in vector notation.
This implies that the projected equations of motion (\ref{dynamics2}) follow from the original ones (\ref{dynamics1}) through the simple substitutions ${\dot{\bf X}}\rightarrow{\dot{\Bf \alpha}}$ and ${\bf F}\rightarrow{\bf f}$.
\section{\label{sec:Appl}  Photon Bose-Einstein Condensate}
In this section we apply the projection optimization method to a concrete problem in the realm of photon Bose-Einstein condensation. Motivated by Ref.~\cite{Wouters2} we derive at first in Subsection \ref{complexGPE} the underlying cGPE in two dimensions for a photon Bose–Einstein condensate \cite{Keeling2008}.  Subsequently,  in Subsection \ref{single}  we formulate and analytically solve 
the projection optimization equation for a homogeneous condensate in the presence of a single vortex steady state.
\subsection{Complex Gross-Pitaevskii Equation}\label{complexGPE}
A photon Bose--Einstein condensate represents a prime example of an open-dissipative system and is therefore predestined for applying the projection optimization method. In the following we consider a photon condensate generated in a dye-filled microcavity \cite{Weitz2010}, where the microscopic physics is well understood \cite{Radonjic,Kirton-Keeling}. Due to multiple absorptions and emissions of photons by the dye molecules, a thermalization is achieved. This is described by a detailed balance condition in form of the Kennard-Stepanov relation \cite{Kennard,Stepanov,Schmitt_Tutorial} between the coefficients of absorption $B_{12} (\omega)$ and emission $B_{21}(\omega)$ according to
\begin{equation}
    \label{Kennard_Stepanov}
    \frac{B_{12}(\omega)}{B_{21}(\omega)} = e^{\hbar\beta\left(\omega-\omega_{\text{ZPL}}\right)}\, ,
\end{equation}
with the inverse temperature $\beta = 1/k_{\rm B}T$, the photon frequency $\omega$, and the zero-phonon-line frequency $\omega_{\text{ZPL}}$. The dynamics of a single-mode system is heuristically motivated at the mean-field level \cite{Schmitt_Tutorial}. To this end the photon field $\Psi$ is modeled by a generalized Gross--Pitaevskii equation (gGPE) \cite{GladilinClassicalField2020phys}
\begin{eqnarray}
    \label{gGPE}
    i\hbar\frac{\partial\Psi}{\partial t} &=& \bigg[-\frac{\hbar^2}{2m}\,\boldsymbol{\nabla}^2 + V({\bf r}) + g\left|\Psi\right|^2 \nonumber \\
    &&+ \frac{i}{2}\left(B_{21}M_2 - B_{12}M_1 - \Sigma \right)\bigg]\Psi,
\end{eqnarray}
where $V$ describes an external potential. Moreover, $m$ and $g$ denote the effective mass and the photon-photon interaction strength, respectively. The additional imaginary term includes the absorption $B_{12}M_1$ and the
induced emission $B_{21}M_2$ of photons due to having $M_1$ and $M_2$ dye molecules in ground and excited state, as well as the losses $\Sigma$ of the empty cavity. Note that the spontaneous emission is neglected in (\ref{gGPE}), as it would result in an additional stochastic noise term. For a complete description of the dynamics, the gGPE (\ref{gGPE}) is self-consistently coupled to rate equations for the number of excited and ground-state molecules \cite{Schmitt_Tutorial}
\begin{eqnarray}
    \label{rate_eq1}
    \frac{\partial M_2}{\partial t} &=& pM_1  - q M_2 - \left(B_{21}M_2 - B_{12}M_1\right)\left|\Psi\right|^2 ,\\
    \label{rate_eq2}
    \frac{\partial M_1}{\partial t} &=& -\frac{\partial M_2}{\partial t}\,,
\end{eqnarray}
where the latter equation follows from the conservation of the total molecule number $M=M_1+M_2$. In (\ref{rate_eq1}), the first term $pM_1$ represents a global pumping term with strength $p$ and the second term $q M_2$ denotes non-radiative decays, while $B_{21}M_2|\Psi|^2$ and $B_{12}M_1|\Psi|^2$ correspond to the induced emission and absorption of photons, respectively.
In order to obtain a simplified description, we introduce the weighted molecule  population inversion $D = B_{21}M_2 - B_{12}M_1$ and consider its temporal change
\begin{eqnarray}
    \label{temp_change_D}
    \dot{D} &=& - \left[ p+q +  \left(B_{21}+B_{12}\right) \left|\Psi\right|^2\right] D \nonumber \\
    &&+ \left( B_{21}p- q B_{12} \right)M\,.
\end{eqnarray}
Note that Eq.~(\ref{temp_change_D}) contains the dynamics of the population inversion of a laser  \cite{Haken1981} in the special case that the coefficients of absorption and emission coincide, i.e.~$B_{21}=B_{12}$. For the considered single-mode system this
formally corresponds to $\omega=\omega_{\rm ZPL}$ according to the Kennard-Stepanov relation (\ref{Kennard_Stepanov}). \\
The steady state of the system is characterized by $\dot{D}=0$ and $\Psi = \Psi_0\, e^{-i \mu t / \hbar}$, where $\mu$ stands for the chemical potential of the photon Bose-Einstein condensate. 
Thus, Eq.~(\ref{temp_change_D}) determines the weighted population inversion
\begin{equation}
    \label{weighted_imbalance}
    D = \frac{D_0}{1+\left|\Psi_0\right|^2 / \Bar{n}}
\end{equation}
where $D_0=(B_{21}p-B_{12}q)M/(p+q)$ denotes the population inversion solely due to the
pump and relaxation processes and $\Bar{n}=(p+q)/\left(B_{12}+B_{21}\right)$ stands for the saturation photon condensate density. 
Merging (\ref{weighted_imbalance}) with (\ref{gGPE}), we obtain the following generalized Gross--Pitaevskii equation for the steady-state photon field
\begin{eqnarray}
    \label{gGPE2}
    \mu \Psi_0 = \Bigg[-\frac{\hbar^2}{2m}\,\boldsymbol{\nabla}^2 + V({\bf r}) + g\left|\Psi_0\right|^2 \nonumber \\
    + \frac{i}{2}\left(\frac{D_0}{1+\left|\Psi_0\right|^2 / \Bar{n}} - \Sigma\right)\Bigg]\Psi_0,
\end{eqnarray}
where the imaginary term on the right-hand side is interpreted as the difference of particle gains and losses. It is worth noting that models similar to (\ref{gGPE2}) are applied in various open-dissipative systems, such as exciton-polariton condensates \cite{RevModPhys.85.299,Bobrovska_exciton}, topological lasers \cite{Carusotto_laser}, and more generally in non-equilibrium quantum fluids \cite{Wouters2}.

In the following, we focus on describing a homogeneous open-dissipative system, i.e.~we put $V(\textbf{r})=0$. In case that the photon number is small enough, Eq.~(\ref{gGPE2}) reduces to the complex Gross-Pitaevskii equation (cGPE) heuristically proposed in Ref.~\cite{Keeling2008}
\begin{eqnarray}
\label{complex}
\hspace*{-5mm} \mu \Psi_0 = \left[ -\frac{\hbar^2}{2m} \,{ \grad }^{2}
+ g \left|\Psi_0\right|^2  + \frac{i}{2}\left( \gamma-\Gamma |\Psi_0|^2 \right) \right] \Psi_0 \,.
\end{eqnarray}
Here $\gamma = D_0 - \Sigma$ stands for an effective pumping strength, whereas $\Gamma |\Psi_0|^2$ 
represents a density-dependent dissipation with
strength $\Gamma = D_0 / \Bar{n}$.
Note that the nonlinear dissipation in (\ref{complex}) can also be understood from
the phenomenological point of view of modeling the dye bleaching as a limiting factor of a single experimental cycle \cite{Weitz2010,Klaers2011,Stein2023a}. And, furthermore, the mean-field model (\ref{complex}) can also be derived microscopically from a Lindblad master equation, where single-particle incoherent pumping and two-body loss are described by corresponding jump operators \cite{Diehl}.
While Ref.~\cite{Keeling2008} investigates (\ref{complex}) for a vortex-free steady-state solution, we determine now the steady-state condensate wave function in the presence of a single vortex.
\subsection{Single Vortex Steady State}
\label{single}
Here we apply the projection optimization method to the case of a single vortex in a homogeneous system and work out an analytical solution.
Assuming that the vortex is located at the origin and has the asymptopic behaviour $\Psi_0({\bf x}) \rightarrow \sqrt{n_{\rm s}}$ for $|{\bf x} | \rightarrow \infty$, this
fixes both the chemical potential $\mu= gn_{\rm s}$ and the saturation density $n_{\rm s}=\gamma/\Gamma$. Furthermore, we use the polar coordinates $r$, $\varphi$ and
decompose the time-independent wave function according to the Madelung transformation
\begin{equation}
    \label{madelung}
    \Psi_0 (r,\varphi) = \sqrt{n_{\rm s}}\,\sqrt{n(r)}\,e^{i\Phi(r,\varphi)}\, ,
\end{equation}
where the dimensionless density profile $n(r)$ is assumed to not depend on $\varphi$ due to cylindrical symmetry.
In addition, we introduce the total velocity field via ${\bf v}(r,\varphi) =\hbar\grad{\Phi}(r,\varphi)/m$ and decompose it using the Helmholtz vector decomposition \cite{Helmholtz} into ${\bf v}(r,\varphi) = {\bf v}_t(\varphi) + {\bf v}_r (r)$ with the rotational tangential part ${\bf v}_t(\varphi)$ and the non-rotational radial part ${\bf v}_r (r)$.
This leads via ${\bf v}_t(\varphi) =\hbar\grad{\Phi_t}(\varphi)/m = \hbar l {\bf e}_\varphi/(mr)$ and ${\bf v}_r(r) =\hbar\grad{\Phi_r}(r)/m$ to corresponding phases, where we assumed in analogy to closed systems that the rotational phase is given by $\Phi_t = l \varphi$ with the vorticity $l=\pm 1$.
By separating the real and imaginary parts of Eq.~\eqref{complex} we obtain two equations
\begin{eqnarray}
    \label{density}
    0 &=& \mu + \frac{\hbar^2}{2m}\left( \frac{\grad^2 \sqrt{n}}{\sqrt{n}} - \frac{l^2}{r^2} \right)- \frac{m}{2} \,{\bf v}_r^2 - gn_{\rm s}\,n\, ,\\
    \label{radialvel}
    0 &=& {\grad}\cdot{\bf v}_r
    + 2 \frac{{\grad} \sqrt{n}}{\sqrt{n}}\cdot
    {\bf v}_r - \frac{1}{\hbar}\left(\gamma-\Gamma n_{\rm s}n\right) \, ,
\end{eqnarray}
which define the density and the radial velocity, respectively. The latter represents an inhomogeneous ordinary differential equation of first order for the radial velocity, which can approximately be solved by choosing a physically reasonable trial ansatz for the density. To this end we read off from (\ref{density}) and (\ref{radialvel}) that the density is only indirectly affected by pump and losses via the radial velocity. Therefore we
assume that the density profile for a vortex in this open-dissipative system coincides with the one in a closed system \cite{Fetter1965,Pitaevskii2016}, i.e.
\begin{equation}
    \label{ansatz}
    n(r) = \frac{r^2}{r^2+\alpha^2}\,,
\end{equation}
where the vortex width $\alpha$ stands for a yet unknown length scale. To determine a physically reasonable value for the trial parameter $\alpha$ one can use the projection optimization method (\ref{variational-projection}) for the Hilbert space $\mathcal{H}=L^2$ with the scalar product (\ref{eq:inner-product}) in $D=2$ dimensions. Inserting therein the equation of motion (\ref{complex}) together with the Madelung transformation (\ref{madelung}), the Helmholtz decomposition for the velocity field and the density profile (\ref{ansatz}) yields
\begin{equation}
    \label{alphaproj}
    \frac{\hbar^2}{2m}\int_0^\infty\dd{r}\frac{r^3}{\left(r^2 + \alpha^2\right)^2}\,{\bf v}_r^2 (r) = \frac{g\gamma}{4\Gamma} - \frac{\hbar^2}{4m\alpha^2}\, .
\end{equation}
This relation connects the radial velocity field ${\bf v}_r$  with the vortex width $\alpha$.
Note that an alternative way to derive Eq.~\eqref{alphaproj} relies on considering  (\ref{density}) as the underlying equation of motion for the density and applying the projection optimization method to it with the density ansatz (\ref{ansatz}).
In order to evaluate  Eq.~\eqref{alphaproj} further, we first need the solution of  Eq.~\eqref{radialvel}. Using (\ref{ansatz}) and applying standard techniques together with the Dirichlet boundary condition that the radial velocity vanishes at the origin, one obtains
\begin{equation}
    \label{radialsolution}
    {\bf v}_r (r)= \frac{\alpha^2\gamma}{2\hbar}\left[\frac{r^2+\alpha^2}{r^3}\,\ln\left(\frac{r^2+\alpha^2}{\alpha^2}\right) - \frac{1}{r}\right] {\bf e}_r\,.
\end{equation}
Thus, the radial velocity turns out to also vanish in the limit $\left|{\bf r}\right|\rightarrow \infty$.
Combining Eqs.~\eqref{alphaproj} and \eqref{radialsolution} yields for the vortex width $\alpha$ two solutions. From these we select  the one, which reproduces in the limit $\Gamma,\gamma\rightarrow 0$ the result $\alpha = \sqrt{2} \xi$ with the coherence length $\xi = \hbar/\sqrt{2mgn_{\rm s}}$, as this is known from the literature for the closed system case \cite{Fetter1965,Pitaevskii2016}.
With this we obtain
\begin{equation}
    \label{alpha}
    \alpha = 2\xi\sqrt{\left({\displaystyle \frac{g}{\Gamma}}\right)^2\left[1 - \sqrt{1-\left(\frac{\Gamma}{g}\right)^2}\,\right]} \, .
\end{equation}
Note that, 
demanding a real-valued density,  Eq.~\eqref{alpha}
 implies a restriction for the losses $\Gamma$ in units of $g$ according to $0 \leq|\Gamma/g| \leq 1$. Due to the relation between the chemical potential $\mu$ and the saturation density $n_{\rm s}$ this also implies a restriction for the pumping $\gamma$ in terms of $\mu$ according to $0 \leq|\gamma/\mu| \leq 1$. In order to have a complete description of the system, 
we conclude this section by mentioning the radial phase, which follows from integrating the radial velocity (\ref{radialsolution})
\begin{eqnarray}
    \label{radial_phase}
    \Phi_r(r)& =& -\frac{\alpha^2\gamma m}{4\hbar^2}\bigg[
    \zeta_2\left(-\frac{r^2}{\alpha^2}\right) - 1
    \nonumber\\
   && + \ln(\frac{r^2+\alpha^2}{\alpha^2})\frac{r^2+\alpha^2}{r^2} 
   \bigg]\, .
\end{eqnarray}
Here $\zeta_2$ stands for the second polylogarithm function and the integration constant is chosen such that the radial phase vanishes at the origin.
\section{Numerical Simulation and Discussion}\label{numerics}
In the following we compare the analytical solution of the cGPE (\ref{complex}) derived in the previous section with a corresponding numerical simulation. To achieve this, we first introduce the numerical method employed. We then examine the impact of finite-size effects on the numerical solution and conclude with a comprehensive comparison of the analytical and numerical steady-state solutions.
\subsection{Numerical Method}
We start with
writing the cGPE (\ref{complex}) in dimensionless form. To this end we scale the particle density with the saturation density $n_{\rm s}$ as well as choose for length and time the appropriate units of $\xi$ and $\xi/\sqrt{2}c_{\rm s}$, respectively, with $c_{\rm s} = \sqrt{gn_{\rm s}/m}$ denoting the sound velocity. With this we end up with
\begin{equation}
\label{eq:cGPEndim}
i \,\frac{\partial \psi}{\partial t} =   \left[ -\grad^{2} + \lvert{\psi}\rvert^{2} - 1 + \frac{i}{2}\,\sigma (1 - \lvert{\psi}\rvert^{2})
    \right]\psi \, ,
\end{equation}
where the remaining dimensionless parameter $\sigma = \Gamma/g$ describes the losses. The dimensionless partial differential equation
\eqref{eq:cGPEndim} is numerically solved by employing the pseudospectral method~\cite{BaoAnExplicitUnc2003siam} with the simulator XMDS2~\cite{DennisXmds2FastSc2013comp}.
To this end, time stepping is performed using a 4th-order adaptive Runge–Kutta integration scheme with a minimum time step of $\Delta_{t} = 10^{-5}$. The spatial part is solved using a cosine basis applied to a box with side lengths $L_{x} = L_{y} = L$ and a grid spacing of $\Delta_{x} = \Delta_{y} = 0.5$. The cosine basis effectively imposes zero Neumann boundary conditions at the box edges.
The initial condition is prepared using
\begin{equation}
\label{initial}
    \psi_{0} (r,\varphi) = \frac{r}{\sqrt{r^{2} + 2}}\,e^{i l \varphi}
\end{equation}
with $r = \sqrt{x^{2} + y^{2}}$, $\varphi = \arctan{(y/x)}$ and $l=1$.
Thus, choosing $x \in [-L_{x}/2, L_{x}/2]$ and $y \in [-L_{y}/2, L_{y}/2]$ puts initially a vortex with constant circulation in the center of the integration box as it is known from a closed system behaviour \cite{Pitaevskii2016}.
For long enough simulation time, which is of the order of
$t_{\rm s} \sim 5\times{}10^{3}$,
we obtain with this a steady-state flow around the vortex, which is noticeably different from the initial condition (\ref{initial}).
\begin{figure}[t]
    \centering
    \includegraphics[width=\columnwidth]{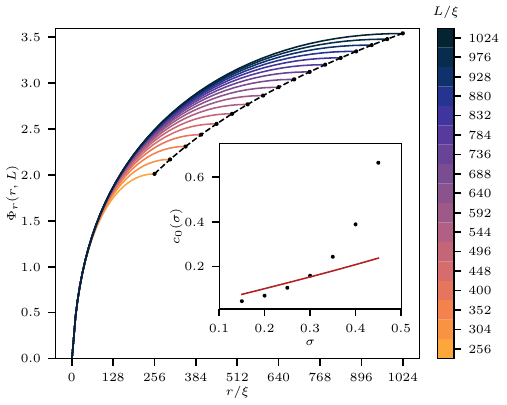}
    \caption{Finite-size effects and radial phase divergence for $\sigma = 0.3$. Each curve shows the radial phase obtained from numerical simulations with varying box sizes $L$. The dotted line represents a second-order polynomial fit in $\ln(L)$, as described by Eq.~\eqref{eq:fitting_func}, illustrating the phase growth with increasing box size. Inset: Leading order coefficient as function of loss parameter $\sigma$ with black dots and red line representing the numerical result and  the expected behaviour extracted from asymptotic analysis of the analytically obtained phase field, see
    Eq.~(\ref{eq:phase_inf_asymptotic}).}
    \label{fig:phase_divergency}
\end{figure}
\subsection{Finite-Size Effects}
Unlike the analytical case, where we have $L = \infty$, all numerical simulations are performed for a finite-sized box of length $L<\infty$. Consequently, it is essential to examine the influence of finite-size effects.
To address this delicate issue, we analyze the numerically determined radial phase for different box sizes $L$, as illustrated in Fig.~\ref{fig:phase_divergency} for $\sigma = 0.3$. Additionally, from the analytically determined radial phase (\ref{radial_phase}) we read off that it diverges logarithmically in the large-distance limit:
\begin{equation}
    \label{eq:phase_inf_asymptotic}
    \Phi_r\left(r,L=\infty\right) \xrightarrow{r \rightarrow \infty} \frac{\alpha^2 \gamma m}{2\hbar^2}\ln^2(r)\,.
\end{equation}
Note that divergent phases are not uncommon in open dissipative systems, as discussed in Refs.~\cite{Aguareles_dyn_cGLE, Aguareles_motion_cGLE, Liu_cGLE} for solutions of the generalized Ginzburg--Landau equation or in Refs.~\cite{Diehl, Diehl_caKPZ} for solutions of the anisotropic Kardar-Parisi-Zhang equation.

In contrast to (\ref{eq:phase_inf_asymptotic}), however, a constant numerical radial phase emerges for every finite box size $L$ due to the effective Neumann boundary conditions. Furthermore, as shown in Fig.~\ref{fig:phase_divergency}, this constant value increases with the box size $L$.
The phase growth is fitted by using a second-order polynomial in $\ln(L)$, which is inspired by the analytical result (\ref{eq:phase_inf_asymptotic}):
\begin{equation}
    \label{eq:fitting_func}
    \Phi_r(L,L) = c_0 \ln^2(L) + c_1 \ln(L) + c_2\,.
\end{equation}
enabling a comparison of the leading-order coefficient. 
As shown in Fig.~\ref{fig:phase_divergency}, the numerically fitted data for $\sigma = 0.3$ shows a very good quantitative agreement with the analytically predicted first-order logarithmic divergence. Moreover, for small loss parameters $\sigma$, the analytically derived and numerically extracted leading-order coefficients align well. However, for $\sigma > 0.3$, the numerical coefficient grows significantly faster with increasing losses compared to the analytical coefficient. This discrepancy implies that, at larger distances, the numerically determined radial phase crosses the analytical result at a certain distance, depending on $\sigma$. This behaviour contrasts with other comparisons between numerical and analytical results, where the analytical values always remain above the numerical ones.
This can be explained as follows: with increasing losses $\sigma$, the `openness' of the system grows, causing the velocity field to reach higher absolute values. As a result, photons accelerate more rapidly than for smaller loss parameters and are influenced by the numerically induced Neumann boundary conditions at much smaller distances. Consequently, the crossing of analytical and numerical values arises due to finite-size effects. For higher $\sigma$, not only does the leading-order divergence in Eq.~(\ref{eq:phase_inf_asymptotic}) play a role in the numerical fitting function (\ref{eq:fitting_func}), but the values of $c_0$ and $c_1$ can no longer be chosen independently, as was the case in the analytical approach.
In addition, while only a qualitative comparison between analytical and numerical results is feasible for larger $\sigma$, the leading-order divergence was successfully predicted analytically and verified through numerical simulations independent on $\sigma$.

Nevertheless, Fig.~\ref{fig:phase_divergency} also shows that for small distances all radial phase profiles coincide regardless of $L$. Therefore, it is reasonable to neglect finite-size effects in the following when comparing results at small distances.
\begin{figure}[t]
    \centering
\includegraphics[width=\columnwidth]{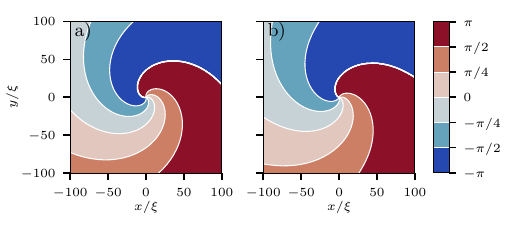}
    \caption{Contour map of the phase around the vortex
    with dimensionless loss parameter $\sigma = 0.40$. The contour lines, shown in white, demonstrate the spiral nature of the vortex. Here a) follows from the projection optimization method, while b) is obtained from the numerical solution.}
    \label{fig:phase_contours}
\end{figure}
\subsection{Comparison with Projection Optimization}
\begin{figure}[t]
    \centering
    \includegraphics[width=\columnwidth]{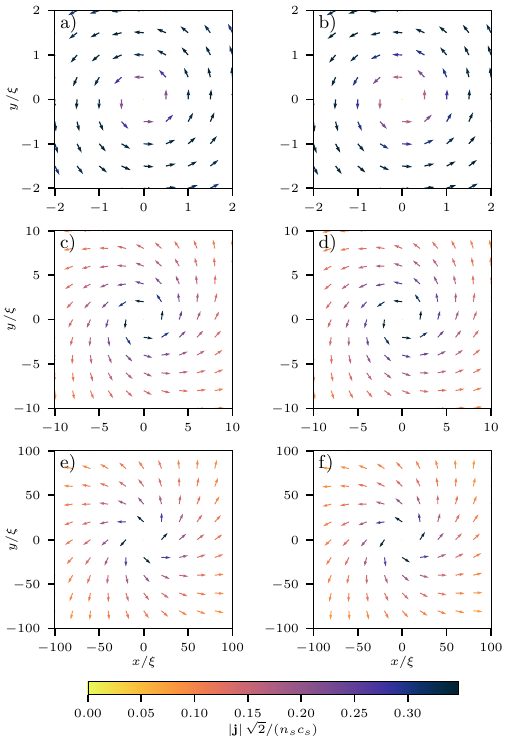}
    \caption{Current around vortex with dimensionless loss parameter $\sigma=0.40$.  Left column (a-c-e) determined analytically using the projection optimization method and right column (b-d-f) obtained by numerically solving cGPE (\ref{eq:cGPEndim}). From top to bottom different characteristics of the flow are visible for varying system length scales. a), b) illustrate that, near the vortex core, the flow is mostly circular similar to the behaviour of vortices in closed system BECs. c), d) depict spiral behaviour at some intermediate distance from the vortex, whereas e), f) show mostly radial behaviour far away from the vortex core.}
    \label{fig:flow_description}
\end{figure}
\begin{figure}[t]
    \centering
\includegraphics[width=\columnwidth]{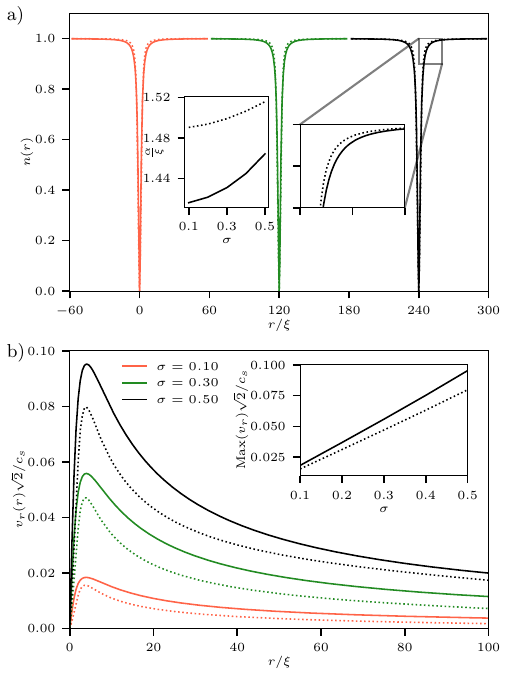}
    \caption{Profiles of a) dimensionsless density, defined in (\ref{madelung}), shifted horizontally for the purpose of illustration, and b) radial velocity for different loss parameters $\sigma$ obtained from  projection optimization method (solid lines) and from solving cGPE (\ref{eq:cGPEndim}) numerically (dashed lines). Insets in a) show
    vortex width $\alpha$ defined in (\ref{ansatz}) and magnified density profile for $\sigma=0.5$. Inset in b)
    depicts that maximal radial velocity changes linearly with the losses.}
    \label{fig:velocity_profile}
\end{figure}

Comparing the results obtained from the
projection optimization method with the corresponding ones from numerics yields at first an excellent agreement. This can be read off, for instance, from the contour map of the phase of the condensate wave function
for the dimensionless loss parameter $\sigma = 0.40$, see Fig.~\ref{fig:phase_contours}.
The contour lines clearly reveal a spiral behaviour of the flow.
This is a direct consequence of a radial velocity contribution ${\bf v}_r$, which results from the open-dissipative nature of the system and competes with the usual tangential velocity field. The nature of the
 competition between circular and radial components of the velocity field can be better visualized by inspecting the current of the condensate flow
${\bf j} =\hbar \Im{\psi\grad{\psi^*}}/m$
with $\Im{ \ldots }$ indicating the imaginary part.
Figure~\ref{fig:flow_description} depicts the flow around a vortex for the same dimensionless loss parameter $\sigma=0.40$ observed at increasing length scales from top to bottom. A visual inspection
seems to indicate  also here that the flow obtained from the projection optimization method agrees well with the corresponding numerical one. This is confirmed by comparing the respective density profiles in
Fig.~\ref{fig:velocity_profile}~a). Increasing the dimensionless loss parameter $\sigma$ yields so tiny differences between the projection optimization method and the numerics, that they are only visible through magnification, see the right inset of Fig.~\ref{fig:velocity_profile}~a). This can also be understood from Eq.~(\ref{density}) as the density is only indirectly affected by the dissipation through the  radial velocity.
However, this turns out to be different for the radial velocity as a function of the distance from the vortex core. From Eq.~(\ref{radialvel}) we recognize that the radial velocity is directly affected by dissipation. And, indeed,
according to Fig.~\ref{fig:velocity_profile}~b) we observe that the projection optimization method leads to a larger deviation from the numerical result for increasing $\sigma$. This is visible, for instance, at the maximal radial velocity, which occurs at the order of the coherence length $\xi$ and increases linearly with the dimensionless loss parameter $\sigma$, see the inset of Fig.~\ref{fig:velocity_profile}~b).
Furthermore,  we read off from Fig.~\ref{fig:velocity_profile}~b) that the radial velocity  falls off slower than $1/r$. Therefore,
the resulting stream lines for the velocity field depicted in Fig.~\ref{fig:flow_description} reveal that one can distinguish three different regions. The stream lines change from being circular in the near field, over spiral at some intermediate distance from the vortex core, up to radial in the far field. This is due to a superposition of an incompressible circular velocity field dominating the near field and a compressible velocity field being relevant in the far field.
This finding from analyzing the vortex solution of the complex Gross--Pitaevskii equation agrees with the numerical work of Ref.~\cite{Wouters2}, where
a nonresonantly excited two-dimensional polariton condensate is analyzed after having adiabatically eliminated the exciton reservoir.

\section{\label{eq:Summary} Summary and Outlook}
We introduced a projection optimization method, which allows an approximative analytic description for open-dissipative quantum fluids.
Applying it to the complex Gross–Pitaevskii equation modeling a photon condensate wave function, we determined the density and phase profiles, along with the corresponding velocity profile for a single vortex. Furthermore, we analytically predicted the divergence of the numerical radial phase for an infinitely large box and demonstrated the finite-size effects.
Furthermore, due to the superposition of an incompressible tangential and a compressible radial component we obtained a spiral vortex shape, which agreed well with the numerics for small dissipation.
In particular, we demonstrated that the projection optimization method is capable of determining the relevant length scale, which characterizes the spiral vortex.
Both the density and the radial velocity turned out to vary on the order of the coherence length.
But whereas for closed systems the coherence length is determined by the density and the interaction strength, here it depends on pump and losses due to the open-dissipative nature of the quantum fluid.
With this we could show exemplarily that the projection optimization method allows to extract useful analytic information for open-dissipative quantum systems at least for small losses, which complements the so far existing numerical studies in the literature.
In order to obtain more accurate analytical results with the projection optimization method in a systematic way one could improve the ansatz for the density (\ref{ansatz}).
To this end one could use the available asymptotic analysis of vortices in closed systems, see e.g.~Ref.~\cite{Berloff_pade,ChenPadeApproximat2021comm}, or one could embark on a corresponding asymptotic analysis for the underlying complex Gross--Pitaevskii equation.
In addition, this work plays a crucial role in understanding the motion of vortex pairs, which, according to established numerical simulations, deviates from the standard point vortex description of closed systems \cite{Wouters1}. Furthermore, we anticipate that these results will have direct implications for our understanding of the Berezinksii--Kosterlitz--Thouless phase transition \cite{berezinskii_1971,berezinskii_1972,Kosterlitz_1973} in open-dissipative systems.

\section*{Acknowledgements}
We are grateful to Enrico Stein for inspiring discussions at an initial stage of the research work. Furthermore, we thank Harry Donegan, Sven Enns, Nikolai Kaschewski,  Kirankumar Karkihalli Umesh, 
Milan Radonji\'c, Julian Schulz, and Georg von Freymann for critically reading the manuscript
 as well as acknowledge financial support by the Deutsche Forschungsgemeinschaft (DFG, German Research Foundation) via the Collaborative Research Center SFB/TR185 (Project No. 277625399). This work was supported by CNPq (Conselho Nacional de Desenvolvimento Cient{\' i}fico e Tecnol{\' o}gico), and DAAD-CAPES PROBRAL, Brazil Grant number 88887.627948/2021-00.

$\mbox{}$\\

\end{document}